\documentclass[conference]{IEEEtran}
\usepackage{fancyhdr}

\usepackage[style=ieee]{biblatex}
\usepackage{amsmath,amssymb,amsfonts}
\usepackage{graphicx}
\usepackage{textcomp}

\usepackage[utf8]{inputenc}
\usepackage[english]{babel}
\usepackage{booktabs}
\usepackage[hidelinks]{hyperref}
\usepackage{nowidow}
\usepackage{bm}
\usepackage[dvipsnames]{xcolor}
\usepackage{mathrsfs}

\usepackage{makecell} 

\newcommand*{\mat}[1]{\ensuremath{\bm{#1}}}
\newcommand*{\sign}{\ensuremath{\text{sign}}}
\newcommand*{\Tr}{\ensuremath{\text{Tr}}}

\begin{document}

\title{Breaking the Exascale Barrier for the Electronic Structure Problem in \textit{Ab-Initio} Molecular Dynamics} 

\author{Robert Schade\IEEEauthorrefmark{1}, Tobias Kenter\IEEEauthorrefmark{1}\IEEEauthorrefmark{2}, Hossam Elgabarty\IEEEauthorrefmark{3}, \IEEEauthorblockN{Michael Lass\IEEEauthorrefmark{1}\IEEEauthorrefmark{2},  Thomas D. Kühne\IEEEauthorrefmark{1}\IEEEauthorrefmark{3}\IEEEauthorrefmark{4}, Christian Plessl\IEEEauthorrefmark{1}\IEEEauthorrefmark{2}}
%
\IEEEauthorblockA{Paderborn University, Warburger Str. 100, 33098 Paderborn, Germany}
\IEEEauthorblockA{\IEEEauthorrefmark{1}Paderborn Center for Parallel Computing \hspace{1ex} \IEEEauthorrefmark{2}Department of Computer Science}
\IEEEauthorblockA{\IEEEauthorrefmark{3}Department of Chemistry \hspace{1ex} \IEEEauthorrefmark{4}Corresponding author, Email: thomas.kuehne@uni-paderborn.de}
}
\maketitle
\thispagestyle{fancy}
\lhead{}
\rhead{}
\chead{}
\rfoot{}
\cfoot{}
\renewcommand{\headrulewidth}{0pt}
\renewcommand{\footrulewidth}{0pt}

\begin{abstract}
The non-orthogonal local submatrix method applied to electronic-structure based molecular dynamics simulations is shown to exceed 1.1 EFLOP/s in FP16/FP32 mixed floating-point arithmetic when using 4,400 NVIDIA A100 GPUs of the Perlmutter system. Thi is enabled by a modification of the original method that pushes the sustained fraction of the peak performance to about 80\%. Example calculations are performed for SARS-CoV-2 spike proteins with up to 83 million atoms. 
\end{abstract}


\section{Introduction}
Electronic-structure based \textit{ab-initio} molecular dynamics simulations (AIMD, \cite{PhysRevLett.55.2471, RevModPhys.64.1045, WIREs}) are an important tool in solid-state physics, chemistry and material science. The explicit treatment of quantum-mechanical effects in the electronic structure is required in situations, where empirical model potentials used in classical molecular dynamics fail to describe the relevant physical or chemical phenomena.

To derive the forces acting on the atoms the electronic-structure problem has to be solved in every time step during the propagation of the atoms. To make this possible, linear-scaling methods have been developed, where the computational complexity scales only linearly with the number of atoms in the system \cite{RevModPhys.71.1085,PhysRevLett.66.1438,PhysRevLett.69.3547,Richters,doi:10.1063/1.4952650}.
We have proposed the non-orthogonal local submatrix method (NOLSM, \cite{SCHADE2022102920}) as a massively parallel method to solve the electronic-structure problem via an approximate solution of the required matrix functions. The local nature of the method avoids inter-node communication in the solution phase and has been shown to scale extremely well to more than one thousand GPUs, while efficiently using the mixed-precision tensor cores for linear algebra operations.

This paper is building on the implementation described in \cite{SCHADE2022102920}, 
but since we are focusing on improvements to increase the sustained peak performance, aspects like the compensation of noise from numerical approximations with an appropriately modified Langevin-type equation to obtain accurate thermodynamical expectation values are not touched here, but have been discussed in previous work \cite{Richters,Computation}. Instead, 
section~\ref{sec:overview} summarizes the tackled problem, whereas Section~\ref{sec:state} puts the achievement in relation to the performance of related large-scale electronic-structure-based structure relaxations or AIMD simulations. The innovations beyond those presented in \cite{SCHADE2022102920} are described in Section~\ref{sec:innovations}. In Section~\ref{sec:measurements}, we discuss our evaluation and define the performance measurements. Finally, Section~\ref{sec:perf} discusses the achieved performance.

\section{Overview of the Problem} \label{sec:overview}
Molecular dynamics calculations simulate the movement of atoms in molecules, surfaces or solids by integrating Newtons equation of motion
\begin{equation}
    M_i \ddot {\mat R}_i=\mat F_i,
    \label{eq:ModLangevinEq}
\end{equation}
where $M_i$ is the mass of atom $i$, ${\mat R}_i$ its position in space and $\mat F_i$ the force acting on it. 
Thus, the evaluation of the forces acting on the atoms is required in every time step. In electronic-structure based molecular dynamics these forces are evaluated on-the-fly from the total energy $E$ of the system via
\begin{equation}
    \mat F_i=-\frac{\partial E}{\partial \mat  R_i}. \label{eq:Eforces}
\end{equation}
In turn, the total energy is not obtained from an empirical model as in classical molecular dynamics, but directly from the quantum-mechanical problem of electrons in the electrostatic field of the nuclei. The total energy of a system can be written as
\begin{equation}\label{eq:etot}
    E=E_{elec} + E_{dc} + E_{ion},
\end{equation}
where $E_{elec}$ is the electronic energy, $E_{dc}$ the double counting terms and $E_{ion}$ the nuclear Coulomb repulsion energy of the atoms.
The bulk of the computational effort is required to obtain the electronic energy. While this can be done efficiently for small and medium-sized systems by solving a high-dimensional eigenvalue problem \cite{PhysRevLett.98.066401,kuhne2020disordered}, very large systems require methods that scale at most linearly with the size of the system in terms of number of atoms. Such linear-scaling electronic-structure methods have been developed \cite{RevModPhys.71.1085,PhysRevLett.66.1438,PhysRevLett.69.3547,Richters} for example based on the one-particle reduced density matrix $\mat D$ \cite{mcweeny1960some}, which allows to obtain the electronic energy via
\begin{equation}\label{eq:e_elec}
    E_{elec}=\Tr(\mat D\mat H_0).
\end{equation}
At zero electronic temperature, the density matrix can be written as a matrix function
\begin{equation}
    \mat D=\frac{1}{2}\left(\mat I-\sign(\mat S^{-1}\mat H_0-\mu \mat I)\right)\mat S^{-1}, \label{eq:purification}
\end{equation} 
where $\mat H_0$ is the one-particle Hamiltonian matrix of the system and $\mat S$ is the overlap matrix between the basis functions that are used to describe the wave functions of the electrons. Furthermore, $\mu$ denotes the chemical potential. 

The evaluation of the matrix-sign function in Eq.~\ref{eq:purification} can be performed iteratively for example with the Newton-Schulz iteration \cite{doi:10.1002/zamm.19330130111}
\begin{align}\label{eq:NS1}
    \mat X_0&=\mat A, \ \ \mat X_{k+1}=\frac{1}{2}\mat X_k (3\mat I-\mat X_k^2)\\ \label{eq:NS2}
    \sign(\mat A)&=\lim_{k\rightarrow \infty} \mat X_k,
\end{align}
or other iteration schemes \cite{Richters_Lass_Walther_Plessl_Kühne_2019} that only require scalings, additions and multiplications of matrices. 

\begin{figure}
    \centering
    \includegraphics[width=0.475\textwidth]{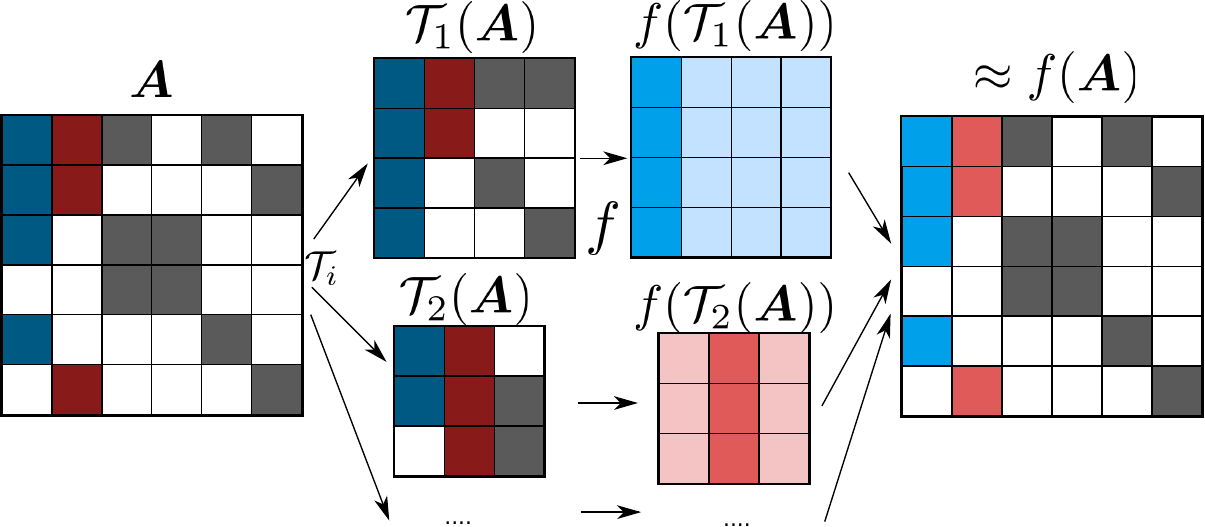}
    \caption{Schematic representation of the steps of the submatrix method for the approximate calculation of a matrix function $f(\mat A)$ of a large sparse matrix $\mat A$. The first step is the construction of a submatrix $\mathcal T_i(\mat A)$ for every column of the matrix $\mat A$. Then the matrix function is applied to the dense submatrices, i.e., $f(\mathcal T_i(A))$ and finally the relevant result columns are inserted into the sparse result matrix. Figure from \cite{SCHADE2022102920}.}
    \label{fig:submatrixgeneral}
\end{figure}

The submatrix method \cite{Lass_Mohr_Wiebeler_Kühne_Plessl_2018, Lass_Schade_Kühne_Plessl_2020} instead views the density matrix as a matrix function to be evaluated. 
Therein, 
the evaluation of a matrix function $f(\mat A)$ is performed in three steps. The steps are performed for every column $i$ of the matrix $\mat A$ independently and are schematically shown in Fig.~\ref{fig:submatrixgeneral}:
\begin{enumerate}
    \item In the first step a submatrix is generated for column $i$ of the input matrix $\mat A$ by removing all rows and corresponding columns for which the matrix $\mat A$ has vanishing or negligible elements in column $i$. The result is a smaller and much denser submatrix $\mathcal T_i(\mat A)$.
    \item The matrix function is applied to the submatrix $\mathcal T_i(\mat A)$, i.e., $f(\mathcal T_i(\mat A))$ is evaluated.
    \item The matrix elements of $f(\mathcal T_i(\mat A))$ that correspond to the column $i$ are written to the matrix $\mat B$ in the sense that the submatrix construction of step 1) is applied in a reverse way.
\end{enumerate}
The resulting matrix $\mat B$ is an approximation of the matrix function $f(\mat A)$ and by construction has the same sparsity pattern as $\mat A$.
The non-orthogonal submatrix method, where $\mat D=\mat D(\mat H_0,\mat S)$ is a matrix function of two matrices, combines the sparsity patterns  $\mat H_0$ and $\mat S$ before the submatrix construction. Thus, it builds two submatrices, $\mathcal T_i(\mat H_0)$ and $\mathcal T_i(\mat S)$ and in step 2) the matrix function is
\begin{equation}
    \mathcal T_i(\mat D)=\frac{1}{2}\left(\mat I-\sign(\mathcal T_i(\mat S)^{-1}\mathcal T_i(\mat H_0)-\mu \mat I)\right)\mathcal T_i(\mat S)^{-1}. \label{eq:purification_submatrix}
\end{equation} 

Please note, that the efficiency and accuracy of the submatrix method can be improved by generating one submatrix for multiple columns instead of just a single column as described in detail in \cite{SCHADE2022102920} together with the GPU implementation. Moreover, 
the use of the submatrix approximation and low-precision numerics can be compensated by an modified Langevin-type equation that replaces Newtons equation of motion so that exact ensemble-averaged expectation values can be obtained \cite{Richters,Computation}.

\section{Current State of the Art} \label{sec:state}
Table \ref{PrevRecords} lists previous attempts to extend the boundaries of electronic-structure based structure relaxation or molecular dynamics simulations.
\begin{table*}[h!]
\centering
\caption{\label{PrevRecords} Performance of previously conducted electronic structure-based structure relaxation or AIMD simulations. Therein, the employed electronic structure method is abbreviated by DFT, NSC-DFT, LS-DFT and SS-DFT, which stands for density functional theory and its non-self-consistent, linear-scaling and subsystem variants, respectively. The corresponding basis set to represent the single-particle orbitals are denoted by PW for conventional plane waves, RMG-PW for real-space multigrid plane waves, GPW for Gaussian and plane waves, GTO for Gaussian-type orbitals, FD for finite difference, RS-FD for real-space finite difference, FEM for finite element method, NGWF for non-orthogonal generalized Wannier functions and PAO for polarized atomic orbitals. If the calculation was conducted involving trivial k-point parallelism, the total number of atoms is given as the product of number of independent instances time the number of atoms in anyone of them. The sustained efficiency is either given with respect to the corresponding peak performance, or estimated in terms of parallel efficiency and identified by the ``$\approx$'' sign. This table has been published previously in \cite{SCHADE2022102920} and is included here with additional results for comparison.}
\begin{tabular}{|l|c|c|c|c|c|c|c|c|c|c|}
     \hline
     \textbf{Code} & \textbf{Year} & \textbf{Method} & \textbf{Basis} & \textbf{System} & \textbf{\# Atoms} & \textbf{\# Cores} & \textbf{Machine} & \textbf{\makecell{Peak\\Performance}} & \textbf{Efficiency} \\ \hline \hline
     CPMD \cite{hutter2005dual} & 2005 & DFT & PW & bulk SiC & 1k & 1.2k CPU & IBM p690 & 1.087 TFLOP/s & $\approx$ 20\% \\ \hline
     Qbox \cite{gygi2006large} & 2006 & DFT & PW & bulk Mo & 8*1k & 128k CPU & IBM BlueGene/L & 207.3 TFLOP/s & 56.5\% \\ \hline
     LS3DF \cite{zhao2009linearly} & 2009 & DFT & PW & bulk ZnTeO & 36k & 147k CPU & Cray Jaguar & 442 TFLOP/s & $\approx$ 33\% \\ \hline
     CONQUEST \cite{bowler2010calculations} & 2010 & NSC-DFT & PAO & bulk Si & 2.1M & 4k CPU & Cray XT4 & & $\approx$ 60\% \\ \hline
     CP2K \cite{vandevondele2012linear} & 2012 & LS-DFT & GPW & bulk H$_2$ & 1M & 47k CPU & Cray XT5 & & \\ \hline
     ONETEP \cite{wilkinson2014hybrid} & 2014 & LS-DFT & NGWF & {\makecell{amyloid fibril\\ trimer}} & 42k & 115k CPU & IBM BlueGene/Q & & \\ \hline
     CONQUEST \cite{arita2014large} & 2014 & LS-DFT & PAO & bulk Si & 786k & 200k CPU & K-Computer & & \\ \hline
     RSDFT \cite{hasegawa2014performance} & 2014 & DFT & RS-FD & Si nanowire & 107k & 664k CPU & K-Computer & 5.48 PFLOP/s & 51.67\% \\ \hline
     CP2K \cite{andermatt2016combining} & 2016 & SS-DFT & GPW & {\makecell{satellite tobacco \\ mosaic virus}}  & 1M & 20k CPU & Cray XC30 & & \\ \hline
     LDC-DFT \cite{nomura2014metascalable} & 2014 & SS-DFT & RMG-PW & bulk SiC & 6.3M & 786k CPU & IBM BlueGene/Q & 5.08 PFLOP/s & 50.5\% \\ \hline
     OpenAtom \cite{jain2016openatom} & 2016 & DFT & PW & periodic MOF & 32*424 & 262k CPU & IBM BlueGene/Q & & $\approx$ 52\% \\ \hline
     MGmol \cite{fattebert2016modeling} & 2016 & LS-DFT & FD & bulk H$_2$O & 1.2M & 1.6m CPU & IBM BlueGene/Q & & $\approx$ 39\% \\ \hline
     DFT-FE \cite{das2019fast} & 2019 & DFT & FEM & Mg cluster & 10.5k & \makecell{159k CPU \\ +22.8k GPUs} & IBM Summit & 46 PFLOP/s & 27.8\% \\ \hline
     CP2K \cite{SCHADE2022102920} & 2021 & LS-DFT & GTO & bulk water & 102M & \makecell{18.4k CPU\\ +1.5k GPUs} & JUWELS Booster & 206 PFLOP/s & 43\% \\ \hline
     CP2K \cite{SCHADE2022102920}  & 2021 & LS-DFT & GTO & {\makecell{HIV-1 capsid \\ in solution}} & 62.5M & \makecell{18.4k CPU \\ +1.5k GPUs} & JUWELS Booster & 324 PFLOP/s & 67.7\% \\ \hline
     CP2K, this work & 2022 & LS-DFT & GTO & {\makecell{spike proteins \\ in solution}} & 82.9M & \makecell{70.4k CPU \\ +4.4k GPUs} & NERSC Perlmutter &  1127 PFLOP/s & 82.1\% \\ \hline
\end{tabular}
\end{table*}

\section{Innovations Realized} \label{sec:innovations}

\subsection{Summary of Contributions}
This work uses the previously reported algorithmic innovations like the use of approximate computing techniques, the non-orthogonal local submatrix method and its realization with GPUs, while minimizing the communication, as well as the heuristic combination of columns in the submatrix creation described in \cite{SCHADE2022102920}.
A new development beyond the implementation innovations already shown in \cite{SCHADE2022102920} like the efficient iterative evaluation of matrix functions for dense matrices on GPU tensor cores is introduced in section~\ref{sec:impl_innovations}: The matrix-size dependency of the GPU-performance is now also considered for the combination of submatrices and yields an additional speedup.

\subsection{Implementation Innovations}\label{sec:impl_innovations}

\subsubsection{Submatrix Combination Heuristics}\label{sec:submatrix_combination}
\begin{figure}
    \centering
    \includegraphics[width=0.475\textwidth]{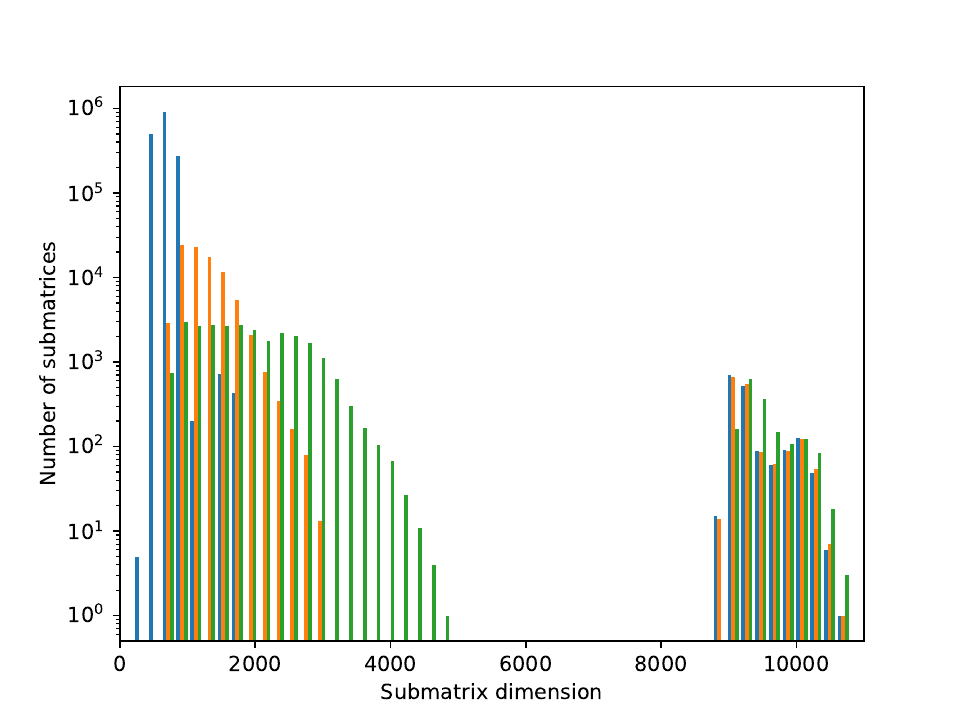}
    \caption{Histogram for the submatrix sizes in the case of the SARS-CoV-2 Spike protein in aqueous solution with approx. 1.7 mio. atoms: without combining submatrices (blue), combination using criterion Eq.~\ref{eq:combination_criterion} (orange) and using criterion Eq.~\ref{eq:combination_criterion2} (green). A discussion of the further structure can be found in \cite{SCHADE2022102920} and also applies to the spike protein system.}
    \label{fig:spike_hist}
\end{figure}

The combination of columns for the generation of submatrices introduced in \cite{SCHADE2022102920} used a cubic metric, i.e., the combination of two columns yields an improvement if and only if 
\begin{equation}\label{eq:combination_criterion}
    {(n_i + n_j - n_{i \land j})}^3 < n_i^3 + n_j^3,
\end{equation}
where $n_i$ is the dimension of the submatrix for column $i$, $n_j$ for column $j$ and $n_{i \land j}$ the number of overlapping columns. This cubic metric represents the number of floating-point operations during the evaluation of the matrix function for the submatrices,  which is based on dense matrix multiplications. 
We propose to modify the cubic metric by including the performance characteristic of the used GPUs. For this purpose, the matrix multiplication performance $p(n)$ of the GPUs is measured for different matrix sizes $n$ and interpolated. The combination criterion then compares the predicted runtime $p(n) \times n^3$ for the matrix functions of the submatrix instead of the number of floating-point operations, i.e.,
\begin{equation}\label{eq:combination_criterion2}
    p(n_i + n_j - n_{i \land j})\times{(n_i + n_j - n_{i \land j})}^3 < p(n_i) \times n_i^3 + p(n_j) \times n_j^3.
\end{equation}
This criterion effectively increases the dimension of the submatrices and the achievable portion of the peak performance. Results are shown for the SARS-CoV-2 Spike protein in aqueous solution with approx. 1.7 mio. atoms in Table~\ref{tab:spike_comb} and histograms for the submatrix dimensions in Fig.~\ref{fig:spike_hist}.
The criterion of Eq.~\ref{eq:combination_criterion2} approximately doubles the average submatrix dimension and slightly increases the total number of floating point operations while drastically increasing the estimated floating-point throughput and leading to an overall estimated speedup compared to the previous criterion of Eq.~\ref{eq:combination_criterion}.

\begin{table}[]
    \centering
    \caption{Influence of the two different submatrix combination criteria, Eq.~\ref{eq:combination_criterion} and Eq.~\ref{eq:combination_criterion2}, on the submatrix sizes, number of floating-point operations for one matrix mutliplication of each submatrix, the estimated performance per NVIDIA A100 GPU and the estimated speedup considering the matrix-multiplication performance of an NVIDIA A100.}
    \label{tab:spike_comb}
    \begin{tabular}{|l|c|c|c|} \hline
        Combination of submatrices & no & Eq.~\ref{eq:combination_criterion} & Eq.~\ref{eq:combination_criterion2} \\ \hline 
        Number of submatrices & 1693134 & 89784 & 28577 \\ \hline
        Smallest submatrix & 396 & 694 & 708 \\  \hline
        Largest submatrix & 10627 & 10637 & 10737 \\ \hline
        Average submatrix dimension & 715 & 1361 & 2282 \\ \hline
        FLOP count in PFLOP for one mult.& 2.02 & 1.54 & 1.66\\ \hline
        Estimated performance in TFLOP/s & 103 & 224 & 270 \\ \hline
        Estimated speedup &  & 2.9 & 3.2 \\ \hline
    \end{tabular}
\end{table}

\section{How Performance Was Measured} \label{sec:perfmeasure}

\subsection{Computational Details}

\subsubsection{SARS-CoV-2 Spike Protein in Aqueous Solution}\label{sec:physical_sys_spike}
\begin{figure}
    \centering
    \includegraphics[width=0.225\textwidth]{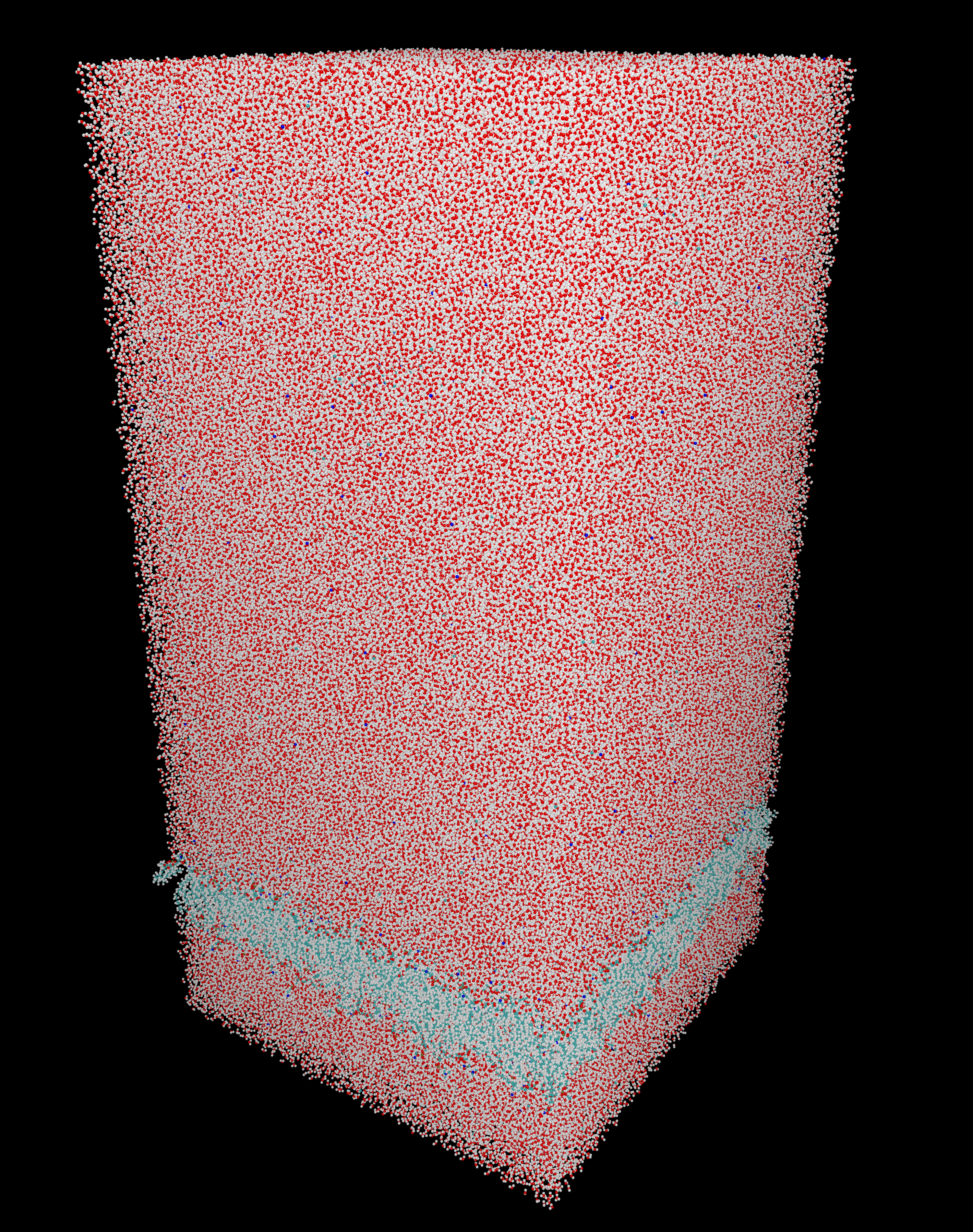}\includegraphics[width=0.225\textwidth]{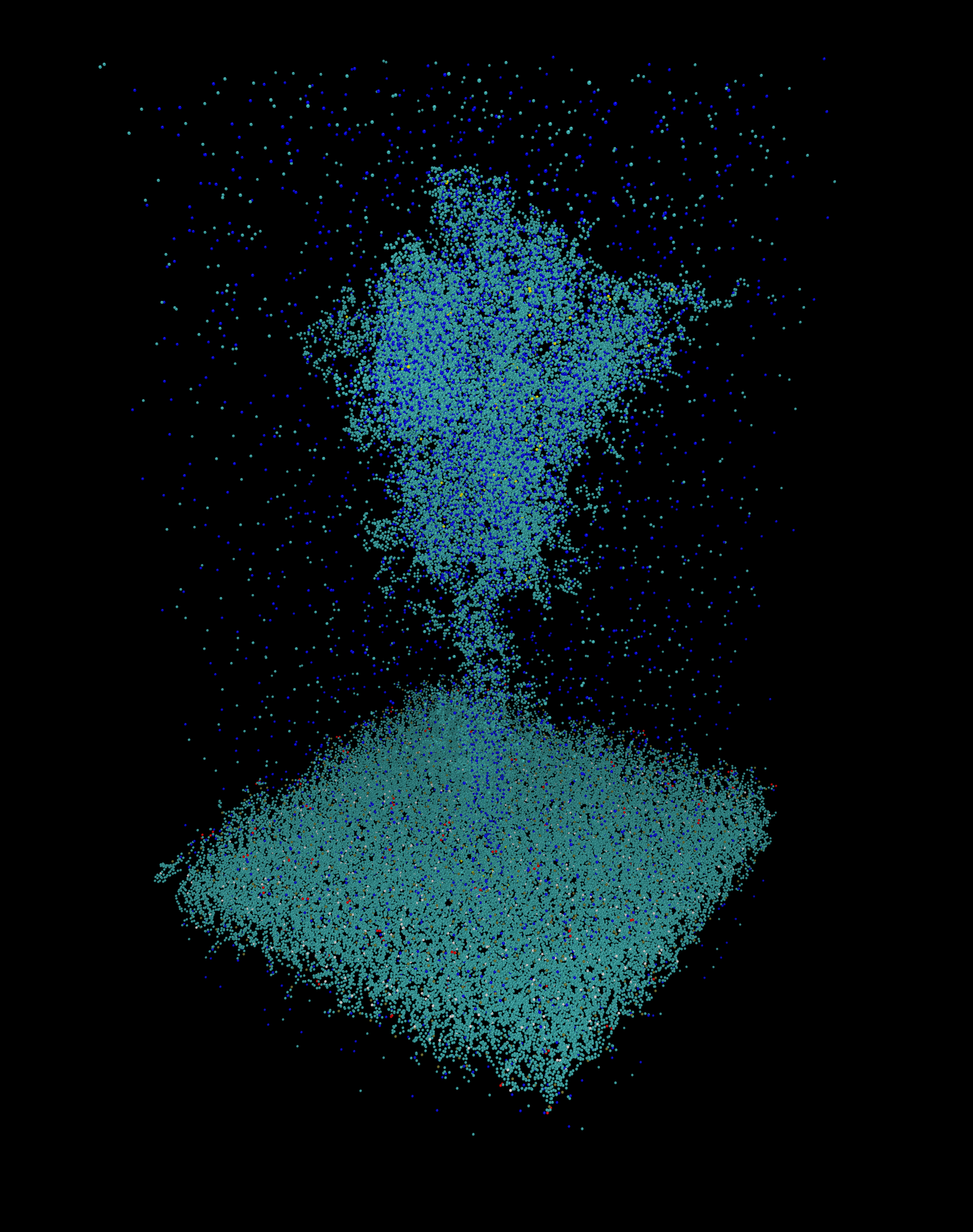}
    \caption{SARS-CoV-2 spike protein in aqueous solution: full cell (left) and without hydrogen and oxygen atoms (right).}
    \label{fig:spike}
\end{figure}

As our benchmark system, we have used the full-length SARS-CoV-2 spike protein in the open state, anchored in a lipid bilayer (Reference PDB structure: 6VSB, and pre-equilibrated with all-atom MD using NAMD \cite{doi:10.1126/science.abb2507,doi:10.1177/10943420211006452}. The system was solvated in aqueous solution in a simulation cell with dimensions 204.7$\times$199.5$\times$408.5~\r{A}, and including 1693134 atoms. The single cell shown in Fig.~\ref{fig:spike} can be easily repeated in a two-dimensional grid of spike proteins as a scalable benchmark system.

\subsubsection{Simulation Details}\label{sec:sim_details}
The electronic structure is simulated with the GFN-xTB approach in conjunction with a London dispersion correction based on the rational Becke–Johnson damping function \cite{grimme2011effect}. Further details can be found in \cite{SCHADE2022102920}.

For the sake of benchmark resources, we have restricted each simulation run to one SCF iteration in the spirit of the second-generation Car-Parrinello AIMD method \cite{PhysRevLett.98.066401,kuhne2018disordered}, but included the iterations for finding an appropriate chemical potential that produces a charge-neutral system. 

\subsection{Measurements}
\label{sec:measurements}
The main measurements presented here are:

\subsubsection{Wall clock time of the NOLSM method $T_{\mathrm{NOLSM}}$}
The wall clock time $T_{\mathrm{NOLSM},i}$ of the NOLSM method on node $i$ is measured for each iteration of the chemical potential. Each iteration includes all transfers between host and GPU. The overall wall clock time  wall clock time $T_{\mathrm{NOLSM}}$ is defined as the maximum over all node wall-clock times.

\subsubsection{FLOPs in the NOLSM method $\mathrm{FLOPs}_{\mathrm{NOLSM},i}$}
\label{sec:NOLSM_Flops}
The per-node floating-point operations $\mathrm{FLOPs}_{\mathrm{NOLSM}}$ in the FP16/FP32-mixed-precision matrix iterations in the NOLSM method are estimated as $2 n^3$ for a gemm-operation $\mat C=\alpha \mat A\cdot \mat B+\beta \mat C$ with $\mat A,\mat B,\mat C \in \mathbb{R}^{n \times n}$ for each iteration of the chemical potential. 
The construction of the matrix elements of the submatrices and other operations scaling like $\mathcal{O}(n^2)$ are neglected in the count.

\subsubsection{Node-Performance of NOLSM method $P_{\mathrm{NOLSM},i}$} 
The node performances of the NOLSM method are defined as $P_{\mathrm{NOLSM},i}=\mathrm{FLOPs}_{\mathrm{NOLSM},i}/T_{\mathrm{NOLSM},i}$ for each node $i$. 

\subsubsection{Performance of NOLSM method $P_{\mathrm{NOLSM}}$} 
The performance of the NOLSM method is defined as the sum of the node performances.

\subsection{HPC System and Environment}
The benchmark runs presented here have been performed on the Perlmutter systsem at the National Energy Research Scientific Computing Center (NERSC). The Perlmutter system consists of 1,536 GPU nodes with one AMD EPYC 7763 64-core CPU with 256 GB DDR4 memory and four NVIDIA A100 GPUs with 40 GB of HBM2 memory each. The peak performance of the tensor cores in one NVIDIA A100 GPU is 312 TFLOP/s in FP16 with FP32-based accumulate~\cite{amperewhitepaper}. The system uses HPE Cray Slingshot as node interconnect.

The software environment used in this work consisted of GCC 11.2.0, Cray-MPICH 8.1.10, CUDA NVCC 11.5.119, and CUBLAS 11.5. 
One MPI-rank per node and 64 CPU-threads per rank were used as well as four CUDA streams per GPU. Each stream was controlled by a single CPU-thread. 

\section{Performance Measurements and Results} \label{sec:perf}

\subsection{Performance of the NOLSM Method for the Spike Protein}
We have performed calculations for three different grid sizes of spike proteins: 6$\times$5 (51 mio. atoms), 6$\times$6 (61 mio. atoms) and 7$\times$7 (83 mio. atoms). All three example calculations have been performed with 1,100 nodes of the Perlmuttter system, i.e., 4,400 NVIDIA A100 GPUs.

\begin{figure}
    \centering
    \includegraphics[width=0.475\textwidth]{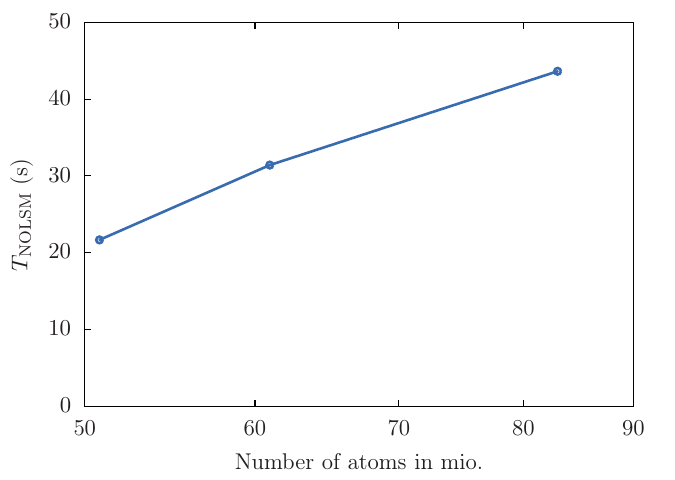}
    \caption{Wall time of the NOLSM method $T_{\mathrm{NOLSM}}$ for a grid of SARS-CoV-2 spike proteins in aqueous solution on 1,100 nodes of the Perlmutter system.}
    \label{fig:spike_wall}
\end{figure}

\begin{figure}
    \centering
    \includegraphics[width=0.475\textwidth]{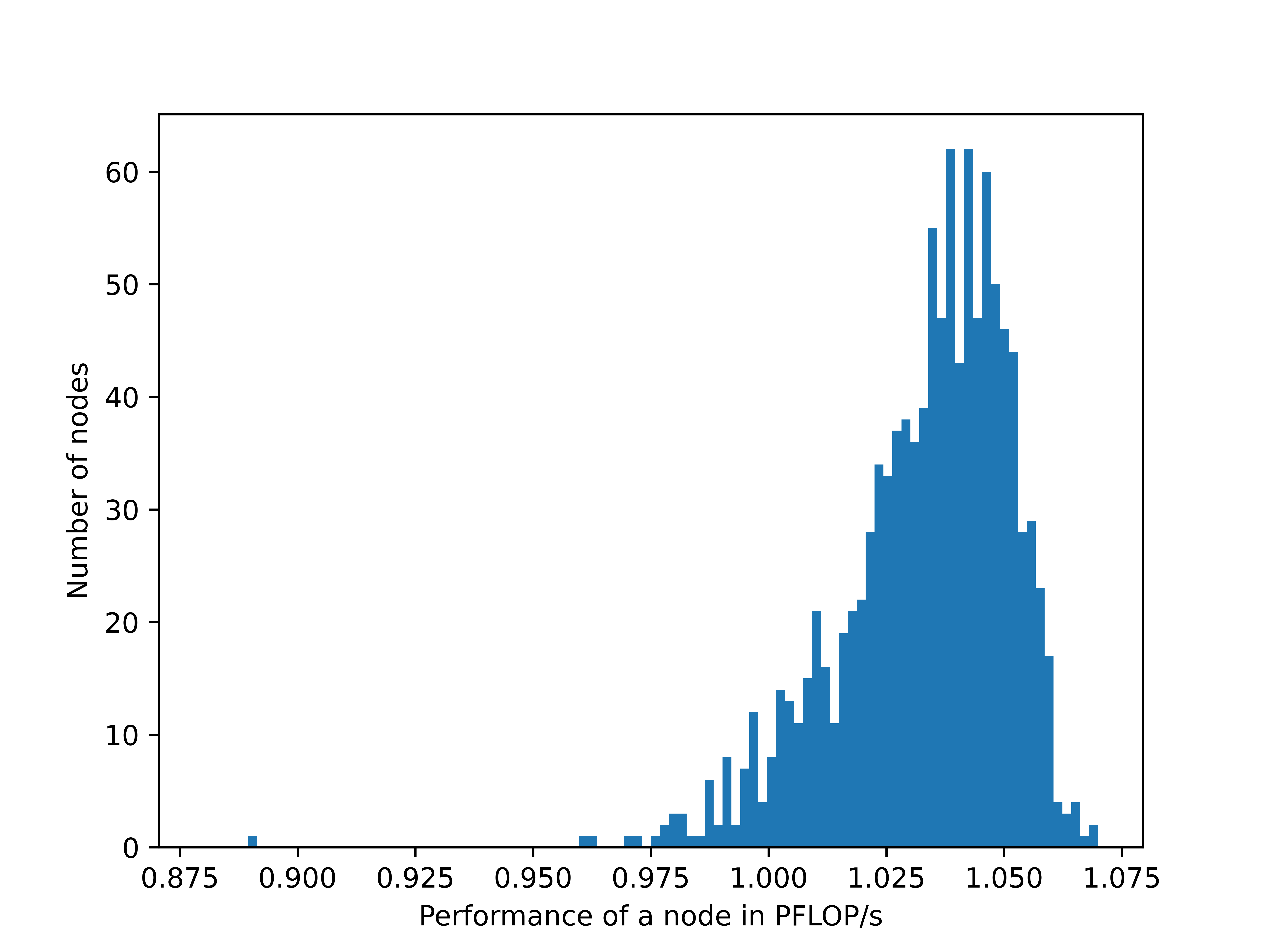}
    \caption{Distribution of node performances for 83 mio. atoms (7$\times$7 grid of SARS-CoV-2 spike proteins in aqueous solution) on 1,100 nodes of the Perlmutter system.}
    \label{fig:spike_perf_hist}
\end{figure}

The wall clock time of the NOLSM method $T_{\mathrm{NOLSM}}$ is shown in Figure~\ref{fig:spike_wall}. 
The distribution of the performances of individual nodes is shown in Figure~\ref{fig:spike_perf_hist} for 7$\times$7 spike proteins (83 mio. atoms) in relation to the peak performance of the GPUs. The performances of the nodes with 4 NVIDIA A100 GPUs mainly fall in the range between 1 PFLOP/s and 1.07 PFLOP/s with an average of 1.03 PFLOP/s. This represents about 80\% of the peak performance of $1.248$ PFLOP/s$=4\cdot 0.312$ PFLOP/s per node.

\begin{figure}
    \centering
    \includegraphics[width=0.475\textwidth]{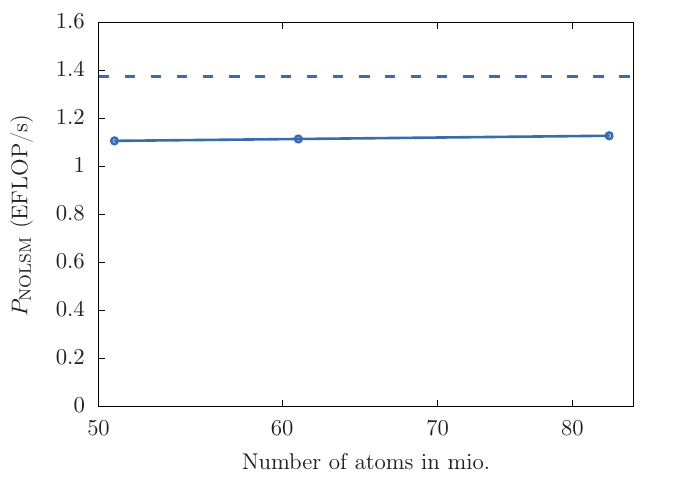}
    \caption{Distribution of node performances for 83 mio. atoms (7$\times$7 grid of SARS-CoV-2 spike proteins in aqueous solution) on 1,100 nodes (4,400 NVIDIA A100 GPUs) of the Perlmutter system.}
    \label{fig:spike_perf}
\end{figure}

Finally, Figure~\ref{fig:spike_perf} shows the floating-point performance $P_{\mathrm{NOLSM}}$ in mixed FP16/FP32 of the NOLSM method. The floating-point throughput of 1.106 to 1.127 EFLOP/s with 4,400 NVIDIA A100 GPUs achieving about 80\% of the theoretical peak performance of the tensor cores. 

\section{Conclusion}
To the best of our knowledge, the achieved $\sim$1.1 EFLOP/s in FP16/FP32 floating-point arithmetic positions electronic-structure based molecular dynamics calculations with the non-orthogonal local submatrix method in CP2K \cite{CP2K} represents one of the first algorithms in computational natural science that has broken the exaflop barrier within a scientific application~\cite{8665799,10.1109/SC.2018.00060,10.1145/3458817.3487399}. 
The massively parallel nature of the method allows for an efficient use of many thousand GPUs. The method can not only be applied to electronic-structure based molecular dynamics, but also in other situations where a matrix function needs to be evaluated for a large sparse matrix or problems that can be transformed to such an operation.

\section*{Acknowledgments}
This research used resources of the National Energy Research Scientific Computing Center (NERSC), a U.S. Department of Energy Office of Science User Facility located at Lawrence Berkeley National Laboratory, operated under Contract No. DE-AC02-05CH11231 using NERSC award DDR-ERCAP0022240.

Additionally, we would like to thank for funding of this project by computing time provided by the Paderborn Center for Parallel Computing (PC2).
This work is partially funded by Paderborn University’s research award for “GreenIT”, as well as
the Federal Ministry of Education and Research (BMBF) and the state of North Rhine-Westphalia as part of the NHR Program.
T.D.K. received funding from the European Research Council (ERC) under the European Union’s Horizon 2020 research and innovation program (Grant Agreement No. 716142).

\printbibliography

\end{document}